\documentclass[aps,amsmath,amssymb,amsfonts,twocolumn,pra,showpacs,nofootinbib]{revtex4}
\usepackage{graphicx}
\usepackage{color}
\usepackage{bm}
\newcommand{\tst}{\textstyle}

\newcommand{\mbf}{\mathbf}
\newcommand{\mrm}{\mathrm}

\begin{document}
\author{Micha{\l} Krych}
\affiliation{Faculty of Physics, University of Warsaw, 00-681 Warsaw, Poland}
\author{Zbigniew Idziaszek}
\affiliation{Institute of Theoretical Physics, Faculty of Physics, University of Warsaw,
00-681 Warsaw, Poland}
\title{Controlled collisions of two ultracold atoms in separate harmonic traps}
\begin{abstract}
We consider controlled collisions between two ultracold atoms guided by external harmonic potentials. We derive analytical solutions of the Schr\"odinger equation for this system, and investigate the properties of eigenergies and eigenstates for different trap geometries as a function of a trap separation and of the scattering length. When varying the trap separation the energy spectrum exhibits avoided crossings, corresponding to trap-induced shape resonances. Introducing an energy-dependent scattering length we investigate the behavior of the system in the vicinity of a magnetic Feshbach resonance. Finally, we illustrate our analytical results with two examples: the quantum phase gate controlled by the external magnetic field, and a scheme for a coherent transport of atoms in optical lattices into higher Bloch bands.
\end{abstract}


\maketitle

\section{Introduction}

Atomic interactions in the presence of an external confinement represent one of the major ingredients for the schemes implementing quantum information processing in systems of neutral trapped atoms (see e.g. \cite{BlochInsight} for a review). The tight external confinement for neutral atoms can be created by means of optical lattices \cite{BlochOptLatt}, atom chips \cite{Dumke,Folman} and dipole traps \cite{Grangier}. Realization of Mott insulator phase \cite{Greiner,StoferleMott,Xu} gives the possibility to precisely control a number of atoms confined in a single well of an optical lattice. Moreover, magnetic Feshbach resonances, widely used in recent experiments on ultracold atoms \cite{Chin}, permit for arbitrary tuning of atomic interactions, which was the key ingredient to accomplish molecular Bose-Einstein condensates \cite{PaulRMP} and the superfluidity in ultracold Fermi gases \cite{Giorgini}. All these achievements make systems of ultracold neutral atoms very attractive in the context of quantum information processing and quantum state control. This has been demonstrated in experiments on creation of massive entanglement \cite{Mandel}, realization of quantum gates \cite{Phillips} or a coherent control \cite{Kohl} of atoms in optical lattices. Recent studies of cold ion-atom collisions \cite{Grier,IdziaszekMQDT} opens also a way for the future realization of quantum computation schemes involving both atoms and ions \cite{IdziaszekIon,Hauke}, combining the advantages of strong interactions of charged ions with long decoherence times of neutral particles.

Modeling of ultracold atomic interactions is most conveniently done in terms of an $s$-wave delta pseudopotential \cite{Fermi,Huang}. In this approach the system of two interacting atoms in a harmonic trap has analytical solutions for spherically symmetric \cite{Busch} or axially symmetric harmonic potentials \cite{IdziaszekPRARapid,IdziaszekPRA}. Moreover, generalization
of the pseudopotential to higher partial waves \cite{StockPseudo,Derevianko,IdziaszekPs} allows to solve the problem for generic types of central interactions in the presence of the harmonic confinement \cite{StockPseudo,IdziaszekPs,Reichenbach,IdziaszekPs1}. The system properties become even more intriguing when the two particles are trapped in separate harmonic potentials. In such a case the system exhibit trap-induced shape resonances \cite{Stock} between molecular and trap states. These resonances, manifesting itself as avoided crossings in an energy spectrum, can be applied, for instance, for a quantum state control \cite{IdziaszekIon,Stock}, or realization of quantum gates \cite{Hauke}.

The setup consisting of two atoms in separate traps can be realized with spin-dependent lattice potentials \cite{BlochSpin}, allowing for individual control of the trapping potential for neutral atoms depending on their internal hyperfine state. Another example is the system of a single atom and a single ion that can be trapped in independent potentials created by optical-dipole and electric radio-frequency fields \cite{IdziaszekIon}. In the latter case the long-range nature of the ion-atom potential prevents from using the contact pseudopotentials to model the interactions, nevertheless, supplementing
the pseudopotential with an energy-dependent scattering length \cite{Blume,Bolda}, allows to consider the regime when the external confinement is of the order of the interaction range. For neutral atoms application of an energy dependent scattering length extends the validity of the
pseudopotential treatment to the case of very tight traps or large scattering lengths in the vicinity of resonances \cite{Bolda2003,IdziaszekPRA}. It also properly accounts for the whole molecular spectrum \cite{Stock}. Discussed models based on the use of pseudopotential provides for a very accurate description of neutral atom interactions in the presence of external confinement. This has been confirmed, for instance, in the recent experiments on the creation of homonuclear \cite{Stoferle} and heteronuclear \cite{Ospelkaus} dimer molecules in optical lattices.

In this paper we present the exact analytical solutions for two interacting atoms confined in separate harmonic potentials. So far the eigenenergies and eigenstates of such a system have been studied only numerically \cite{Stock,Reichenbach,IdziaszekIon}. For simplicity we assume that atoms are confined in the traps of the same trapping frequency, which allows to separate the center-of-mass and relative motions. The numerical studies of the more general situation with different trapping frequencies \cite{IdziaszekIon} show, however, that the basic features consisting in the presence of the trap-induced resonances remain unchanged. We present the analytical results for the energy spectrum and wave functions discussing different geometries of the harmonic trapping potential. Applying the energy-dependent scattering length we investigate the effects of Feshbach resonances on the trap-induced resonances. We illustrate our analysis with two example applications of the trap-induced shape resonances for the atoms in optical lattices. We consider a simple scheme for a quantum phase gate between atoms in separate wells, and we present a method for a coherent transport of atoms into higher Bloch bands.

The paper is organized as follows. In section~\ref{Sec:Model} we derive analytical solutions of the Schr\"odinger equation for two interacting atoms in separate traps. Section~\ref{Sec:Energ} is devoted to the analysis of the energy spectrum. The wave functions are investigated in section~\ref{Sec:Wave}. The energy spectrum in the vicinity of a magnetic Feshbach resonance is analyzed in section~\ref{Sec:Feshbach}. We illustrate the applicability of our analytical results in section~\ref{Sec:Examples} considering the quantum phase gate and the scheme for a coherent transfer between quantum states in the trap. Section~\ref{Sec:Conclusions} present some conclusions, and three appendices give some more technical details on our derivation.

\section{Model}
\label{Sec:Model}

\begin{figure}
\includegraphics[width=8.6cm,clip]{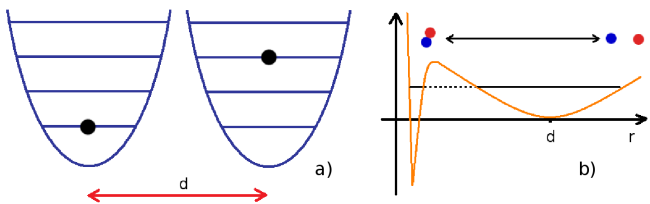}
\caption{
\label{Fig1} (Color online)
a) Two atoms in cylindrical harmonic traps separated by the distance $d$. b) A relative part of the effective potential.
}
\end{figure}
We consider the system of two interacting particles confined in separate harmonic potentials. The setup is illustrated schematically in Fig.~\ref{Fig1}.
We assume that the trapping potentials of the particles can be controlled independently and the distance between traps can be arbitrarily adjusted. The Hamiltonian of the system can be written down as:
\begin{equation}
\begin{split}
  \hat{H}=&-\frac{\hbar^2}{2m_1}\nabla_1^2-\frac{\hbar^2}{2m_2}\nabla_2^2  \\ &+V_{t}^{(1)} (\mbf{r}_1-\mbf{d}_1)+V_{t}^{(2)} (\mbf{r}_2-\mbf{d}_2)+V_i (\mbf{r}_1-\mbf{r}_2).
\end{split}
\end{equation}
where $m_1$, $m_2$ are masses of atoms, $V_t^{(1)}(\mathbf{r})$, $V_t^{(2)}(\mathbf{r})$ are their respective trapping potentials, $\mbf{d}_{1}$, $\mbf{d}_{2}$ denote trap positions, and $V_i(\mathbf{r})$ is the interaction between particles. For simplicity we assume that the harmonic traps are cylindrically symmetric, however, our method to derive analytical solutions is general and can be applied for arbitrary harmonic potential. The trapping potential can be written as
\begin{equation}
V_{t}^{(j)} (\mathbf{r})=\frac{m_j}{2}(\omega_\perp \rho^2 +\omega_z z^2), \quad j=1,2.
\end{equation}
In a harmonic external potential the center-of-mass and relative motions can be separated, and in the following we will focus only on the relative motion that is governed by
\begin{equation}
   \hat{H}_{rel}=-\frac{\hbar^2}{2\mu}\nabla^2+\frac{1}{2}\mu \omega^2 \mathbf{(r-d)^2}+V(\mathbf{r}).
\end{equation}
Here, $\mathbf{r=r_1-r_2}$, denotes the relative coordinate, $\mu=m_1 m_2/(m_1+m_2)$ is the reduced mass, and $\mbf{d}= \mbf{d}_1 - \mbf{d}_2$ is the distance between the traps. Without loosing generality we can assume that traps are displaced along $z$-direction $\mbf{d} =d \hat{\mbf{z}}$. In our model we approximate the atom-atom interaction with a contact Fermi pseudopotential
\begin{equation}
  V(\mathbf{r})=\frac{2 \pi \hbar^2 a}{\mu}\delta(\mathbf{r})\frac{\partial}{\partial r}r
\end{equation}
describing $s$-wave scattering, which dominates at ultracold temperatures. This approximation is valid provided that the characteristic length of the trapping potential is much larger than the characteristic radius of the interaction. This is typically well fulfilled for neutral atom collisions, however it typically breaks down for long-range potentials, like the atom-ion one. In such cases one can apply the energy-dependent scattering length to extend the validity of the pseudopotential to the regime when the trap size is of the order of the interaction range.

In the following we will adopt dimensionless units expressed in terms of the oscillator length $l=\sqrt{\hbar/(\mu \omega_z)}$ and energy $\hbar \omega$. Rewriting the Hamiltonian in cylindrical coordinates yields
\begin{equation}
\label{eq:schr}
\left[-{\tst\frac{1}{2}}\Delta+{\tst\frac{1}{2}}(\eta^2\rho^2+(z-d)^2)+2\pi a\delta(\mathbf{r})\frac{\partial}{\partial r}r\right]\!\!\Psi(\mathbf{r})
\! = \!E\Psi(\mathbf{r}),
\end{equation}
where $\eta=\omega_\perp/\omega_{z}$ is the trap aspect ratio.

\section{Energy spectrum}
\label{Sec:Energ}

We turn now to evaluation of eigenenergies. To this end we decompose $\Psi(\mathbf{r})$ in the basis of the harmonic oscillator wave functions in cylindrical coordinates. Since the delta pseudopotential acts only on states that do not vanish at $\mbf{r} = 0$, in the expansion we include only the harmonic oscillator states with zero projection of the angular momentum on the symmetry axis ($m=0$). The expansion reads
\begin{equation}
\label{eq:exp}
\Psi(\mathbf{r})=\sum_{n,k}c_{n,k}\Phi_{n,0}(\rho,\varphi)\Theta_k(z),
\end{equation}
where $\Phi_{n,0}(\rho,\varphi)$ and $\Theta_k(z)$ are the harmonic oscillator states for one and two dimensions, respectively, in the trapping potential $V_t(\mbf{r})$. Substituting \eqref{eq:exp} into the Schr\"odinger equation \eqref{eq:schr}, and projecting on a single state of the basis gives the expansion coefficients
\begin{equation}
  c_{n,k}=C\frac{\Phi_{n,0}^{*}(0,\varphi)\Theta_k^{*}(-d)}{E-E_{n,k}} \label{eq2},
\end{equation}
where $E_{n,k}=1/2+\eta+k+2\eta n$ are eigenenergies for a cylindrically symmetric trap and
\begin{equation}
  C=2\pi a \left[\frac{\partial}{\partial r}(r \sum_{n',k,}c_{n',k,}\Phi_{n',0}(\rho,\varphi),\Theta_k(z-d))\right]_{r=0}
  \label{eq3}
\end{equation}
is some constant prefactor, given by the normalization of the wave function. Substituting Eq.~(\ref{eq2}) into (\ref{eq3}) we obtain equation determining the energy spectrum
\begin{equation}
\begin{split}
\label{spectr1}
-\frac{1}{2 \pi a}=\left[\frac{\partial}{\partial r}r \Psi_{\epsilon}(\mathbf{r})\right]_{r=0},
\end{split}
\end{equation}
where $\Psi_{\epsilon}(\mathbf{r})$ is an eigenstate corresponding to the eigen\-energy $\epsilon$
\begin{equation}
\Psi_{\epsilon}(\mathbf{r})=\sum_{n,k}\frac{\Phi_{n,0}^{*}(0,\varphi)\Phi_{n,0}(\rho,\varphi)\Theta_k^{*}(-d)\Theta_k(z-d)}{2 \eta n+k-\epsilon}.
\label{falowa}
\end{equation}
Here, $\epsilon=E-E_0$ and $E_0 = 1/2+\eta$ is the energy of the zero-point oscillations. We note that the eigenstate \eqref{falowa} is not normalized. We further proceed by expressing the denominator in Eq.~\eqref{falowa} in terms of an integral
\begin{equation}
  \frac{1}{2\eta n+k+\epsilon}=\int_0^{\infty}dt e^{-t(2\eta n+k+\epsilon)},
  \label{tworz}
\end{equation}
which allows to perform summations over excitation quantum numbers $k$ and $n$. To this end we substitute the explicit formulas for the oscillator wave functions  $\Phi_{n,0}(\rho,\varphi)$ and $\Theta_k(z)$, and utilise the summation formulas for the Hermite and Laguerre polynomials (see Appendix \ref{App:Sums} for details). This yields an integral representation of the wave function
\begin{widetext}
\begin{equation}
\Psi_{\epsilon}(\mathbf{r})=\frac{\eta}{(2\pi)^\frac{3}{2}}\int_0^{\infty}dt\frac{e^{(t\epsilon -\frac{d^2+(z-d)^2}{2}\coth(t)-\frac{\eta \rho^2}{2}\coth(\eta t)+\frac{ d(d-z)}{\sinh t})}}{\sqrt{\sinh t \sinh(\eta t)}}.
\label{IntRepr}
\end{equation}
\end{widetext}
Similar integral representation can be derived for the equation determining the energy spectrum. To this end we substitute \eqref{IntRepr} into \eqref{spectr1}, obtaining
\begin{equation}
-\frac{\sqrt{\pi}}{a}={\cal F}\left(-\frac{\epsilon}{2}\right) = 2 \pi^{3/2} \left[\frac{\partial}{\partial r}r \Psi_{\epsilon}(\mathbf{r})\right]_{r=0},
\label{energie}
\end{equation}
where
\begin{equation}
{\cal F}(x)\equiv\int_0^{\infty}dt{(\frac{\eta e^{-x t} e^{-d^2 \frac{1-e^{-t/2}}{1+e^{-t/2}}}}{(1-e^{-t \eta})\sqrt{1-e^{-t}}}-\frac{1}{t^{3/2}})}.
\label{Fdef}
\end{equation}
The term $t^{-3/2}$ assures that the integral converges at the lower limit of integration, and it results from the application of the regularization operator in the Fermi pseudopotential. As can be easily shown, the wave function \eqref{falowa} exhibits $1/r$ divergent behavior at small $r$, originating from the behavior of the integral \eqref{IntRepr} at $t \rightarrow 0$. Factoring out the divergent term from the integral representation \eqref{IntRepr} in a way described in Ref.~\cite{IdziaszekPRA}, and then removing the divergent $1/r$ term with the help of the regularization operator results in the integral \eqref{Fdef}. In the simplest case of an isotropic trapping potential ($\eta =1$), the integral \eqref{Fdef} can be evaluated in terms of a series
\begin{multline}
   {\cal F}(x,\eta=1) = -2 \sqrt{\pi} \frac{\Gamma(x)}{\Gamma(x-\frac12)} \\
    + \sum_{k=1}^{\infty}\frac{(-d^2)^k}{2 k!} B(2x,k-\tst{\frac{1}{2}}){}_2F_1 (2x-2,k-\tst{\frac{1}{2}};2x+k-\tst{\frac{1}{2}};-1),
\label{roznica d}
\end{multline}
where $B(x,y) = \Gamma(x) \Gamma(y)/\Gamma(x+y)$ denotes the Euler Beta function \cite{Gradshteyn}. For $d=0$ this straightforwardly reduces to the known result for two atoms in isotropic trap \cite{Busch}. In general case the integral \eqref{Fdef} is too complicated for an analytical treatment. While for $x>0$ the integral is well defined and can be calculated numerically, for $x<0$ the integral is defined only in the sense of the analytic continuation. Therefore, in the following we will develop the recursion formula, which relates ${\cal F}(x)$ at $x<0$ with the values of ${\cal F}(x)$ at $x>0$.

\subsection{Recurrence relations}

Let's consider the difference between values of ${\cal F}(x)$ separated by $\eta$
\begin{equation}
{\cal F}(x)-{\cal F}(x+\eta)=\int_0^{\infty}dt{\frac{\eta e^{-x t} e^{-d^2 \frac{1-e^{-t/2}}{1+e^{-t/2}}}}{\sqrt{1-e^{-t}}}}.
\end{equation}
The above integral can be calculated analytically (see Appendix B for details)
\begin{equation}
{\cal F}(x)-{\cal F}(x+\eta)=B\left(\textstyle{\frac12},2x\right) \Phi_1\left(\textstyle{\frac12},2x,2x+\textstyle{\frac12};-1,-d^2\right),
\label{Recur}
\end{equation}
where $\Phi_1(a,b,c;z,w)$ is the degenerate hypergeometric function of the two variables. It is defined as \cite{Gradshteyn}
\begin{equation}
\Phi_1(a,b,c;z,w)=\sum_{m,n=0}^{\infty}\frac{(a)_{m+n} (b)_n}{(c)_{m+n} m! n!}z^m w^n,
\end{equation}
where $(a)_n$ is the Pochhammer symbol. Although the integral representation \eqref{Fdef} is valid for $x>0$, we note that the final result involving $\Phi_1(a,b,c;z,w)$ is well defined for both $x<0$ and $x>0$, by the virtue of the analytic continuation. While $\Phi_1$ allows to express the result of integration in quite elegant form, its numerical evaluation is rather cumbersome. In fact, in the numerical calculations we have applied the following series expansions
\begin{align}
& B\left(\textstyle{\frac12},2x\right) \Phi_1\left(\textstyle{\frac12},2x,2x+\textstyle{\frac12};-1,-d^2\right) \nonumber \\
&=\sum_{n=0}^{\infty} \Bigg[ (-1)^n \frac{\Gamma(n+2x)}{\Gamma(2x) n!}
\nonumber \\
& \qquad \times B(2x,n+\textstyle{\frac12}){{}_1 F_1}(n+\textstyle{\frac12};2x+n+\textstyle{\frac12};-d^2)\Bigg] \\
  &=\frac{2^{2x-1}}{\Gamma(2x)}\sum_{n=0}^{\infty} 2^n\frac{(\Gamma(n+2x))^3}{n! \Gamma(2n+4x)}{{}_1 F_1}(\frac{1}{2};2x+n+\frac{1}{2};-d^2),
\end{align}
which are derived in the Appendix~\ref{App:Recur}.

\subsection{Results}

\begin{figure}
\includegraphics[width=8.6cm,clip]{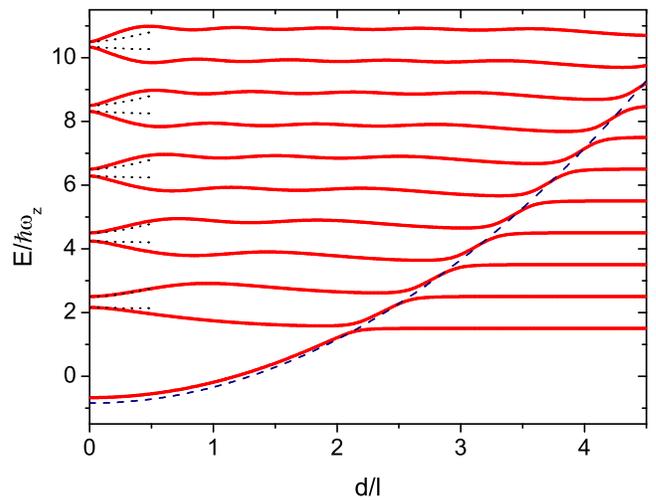}
\caption{
\label{Fig2} (Color online)
Energy spectrum (solid red lines) versus trap separation $d$ for two interacting atoms confined in harmonic traps of geometry $\eta=\omega_\perp/\omega_z=1$, and for $a\approx 0.79 l$. All lengths are expressed in oscillator units: $l = \sqrt{\hbar/(\mu \omega_z)}$. Blue dashed line shows the approximate quadratic shift of the bound state due to the external trapping potential.
Black dotted line is a result of a perturbation calculations between exact spherical zero-separation wave functions and oscillator $p$ states.
}
\end{figure}

We calculate the energy spectrum from Eq.~\eqref{energie} with ${\cal F}(x)$ evaluated numerically from the integral \eqref{Fdef} for $x>0$ and using the recurrence relation \eqref{Recur} for $x<0$. Fig.~\ref{Fig2} shows the results for the spherically symmetric traps and a fixed value of the scattering length, but with varying the trap separation $d$. We observe that for large $d$ the energy levels acquire values of a noninteracting harmonic oscillator. For smaller trap separations we observe avoided crossings with a bound state of the interparticle potential, which is lifted up by the external trapping potential, with approximate quadratic dependence on $d$. This behavior can be explained by observing that the bound state wave function is concentrated around $\mbf{r}=0$, so $\langle \Psi_{mol}(d)|H_{rel}(d)|\Psi_{mol}(d)\rangle \approx E_b+\frac12 \mu\omega_z^2 d^2$, where $E_b$ denotes the binding energy.

The avoided crossings correspond to the trap-induced shape resonances \cite{Stock}, that occur when the energy of bound states lifted by the trapping potential coincide with the energy of a vibrational state in the trap (see Fig~\ref{Fig1}.b). For the spherically symmetric traps ($\eta=1$) at zero separation ($d=0$) the angular momentum of the relative motion is conserved and the states can be additionally characterized by the partial wave quantum number $l$. In Fig.~\ref{Fig2} we observe that at $d=0$ the spectrum contains not only the $s$-wave eigenstates with energies shifted by the zero-range potential \cite{Busch}, but also $p$-wave harmonic oscillator states with energies $E_n = \hbar \omega (5/2 + 2 n)$. The presence of the $p$-wave states in the spectrum can be explained in the framework of the perturbation theory. At small $d$ the external potentials $\frac12 \mu \omega^2(x^2 + y^2 + (z-d)^2)$ couples $s$-wave eigenstates of the interacting atoms with $p$-wave harmonic oscillator states (see Fig.~\ref{Fig2}).

\begin{figure}
\includegraphics[width=8.6cm,clip]{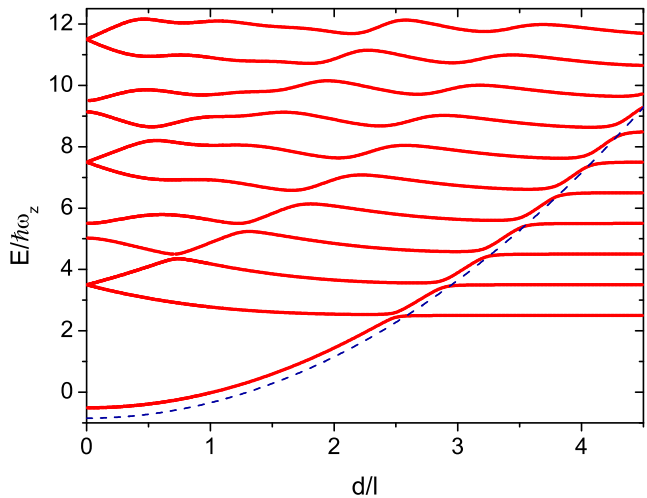}
\caption{
\label{Fig3} (Color online)
Energy spectrum (solid red lines) versus trap separation $d$ for two interacting atoms confined in harmonic traps of geometry $\eta=\omega_\perp/\omega_z=2$, and for $a\approx 0.79 l$. All lengths are expressed in oscillator units: $l = \sqrt{\hbar/(\mu \omega_z)}$. Blue dashed line shows the approximate quadratic shift of the bound state due to the external trapping potential}
\end{figure}

Similar features at small $d$ can be also observed for an anisotropic trap (see Fig.~\ref{Fig3}). In this case eigenstates of the $s$-wave contact potential are coupled in the first order in $d$ to the harmonic oscillator states. In the cylindrically symmetric trap their energies are given by
$E=\hbar\omega(\eta+ 1/2+ n_z+ 2\eta n_\perp)$. By studying the case of non-commensurable trapping frequencies $\omega_z$ and $\omega_\perp$, we have verified that only states with even $n_z$ are present in the spectrum at $d \rightarrow 0$.

\begin{figure}
\includegraphics[width=8.6cm,clip]{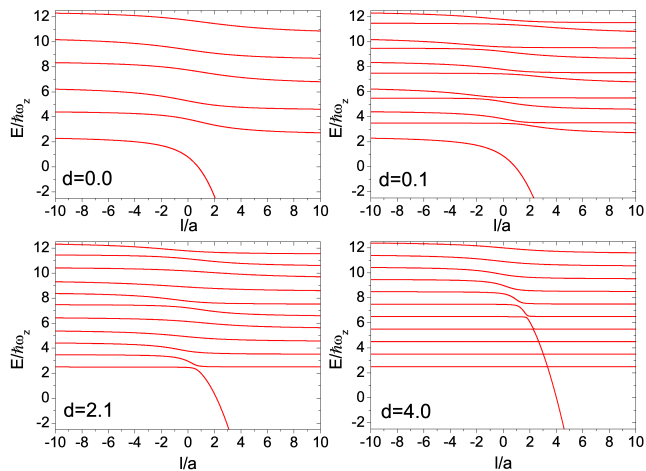}
\caption{
\label{Fig4} (Color online)
Energy spectrum as a function of the inverse scattering length for two interacting atoms confined in harmonic traps of geometry $\eta=\omega_\perp/\omega_z=2$, and for trap separations $d/l=0.0, 0.1, 2.1, 4.0$. All lengths are expressed in oscillator units: $l = \sqrt{\hbar/(\mu \omega_z)}$.}
\end{figure}

\begin{figure}
\includegraphics[width=8.6cm,clip]{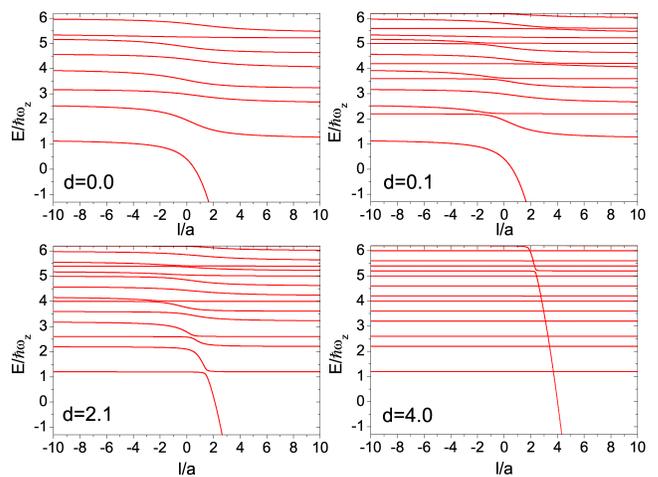}
\caption{
\label{Fig5} (Color online)
Energy spectrum as a function of the inverse scattering length for two interacting atoms confined in harmonic traps of geometry $\eta=\omega_\perp/\omega_z=0.7$, and for trap separations $d/l=0.0, 0.1, 2.1, 4.0$. All lengths are expressed in oscillator units: $l = \sqrt{\hbar/(\mu \omega_z)}$.}
\end{figure}

Figs.~\ref{Fig4} and \ref{Fig5} present the energy spectrum versus the scattering length, for a fixed trap separation, and for two different trap geometries: $\eta=2$ and $\eta=0.7$, respectively. For $d=0$ the spectrum coincides with the one obtained for two trapped atoms in an anisotropic harmonic trap \cite{IdziaszekPRA}. For non-commensurable trapping frequencies $\eta=0.7$ and $d=0$, we note the appearance of some additional states in the spectrum, weakly dependent on $a$. These states give rise to avoided crossings at $a \rightarrow 0$ \cite{IdziaszekPRA}.
For small and positive $d$ the energy spectrum contains additionally states of the harmonic oscillator with even $n_z$, which may be observed as horizontal lines, due to their weak dependence on $a$. They create avoided crossings with the rest of the states that are strongly influenced by the presence of the atom-atom interaction, and which are already present at $d=0$. With increasing $d$, the energy of the bound state rises, and at some moderate trap separation ($d \sim l$), the bound state starts to create avoided crossings with vibrational states in the trap.

\subsection{Energy-dependent scattering length}

So far we have performed our analysis using the standard, energy-independent scattering length, defined in the zero-energy limit: $a = - \lim_{k \rightarrow 0} \tan \delta_0 (k)/k$ in terms of the $s$-wave phase shift $\delta_0(k)$ and the wave vector $k = \sqrt{2 \mu E}/\hbar$. This approach is valid for sufficiently weak trapping potentials: $\hbar \omega \ll \hbar^2/(\mu |a| R_\mrm{eff})$.
In this case the effective range correction in the expansion of the $s$-wave phase shift can be neglected: $k \cot \delta_0(k) = -1/a + \frac12 R_\mrm{eff} k^2 + O(k^3)$, where we can take $\hbar^2 k^2/(2 \mu) \sim \hbar \omega$. Here $R_\mrm{eff}$ denote the effective range, which for a van der Waals interaction is of the order of the characteristic length of the $C_6/r^6$ potential: $R_\mrm{eff} \sim R_\mrm{vdW} \equiv \frac12 (2 \mu C_6/\hbar^2)^{1/4}$ \cite{Gao1998}.

In the situation when the center-of-mass and relative motions are not coupled, the more general treatment of atom-atom interactions is based on the use of an energy-dependent scattering length $a(k) = - \tan \delta_0 (k)/k$ \cite{Blume,Bolda}. This is particularly important for large values of the scattering length
obtained in the presence of scattering resonances, that may occur due to the interchannel coupling (Feshbach resonances) or due to the potential barrier (shape resonances). Moreover, the energy dependent scattering length extends the validity of the pseudopotential to the case of tight traps.
In such system the standard, energy-independent pseudopotential becomes inaccurate because of the relatively high kinetic energy of the colliding atoms and comparable range of the interaction potential and of the external confinement.

\begin{figure}
\includegraphics[width=8.6cm,clip]{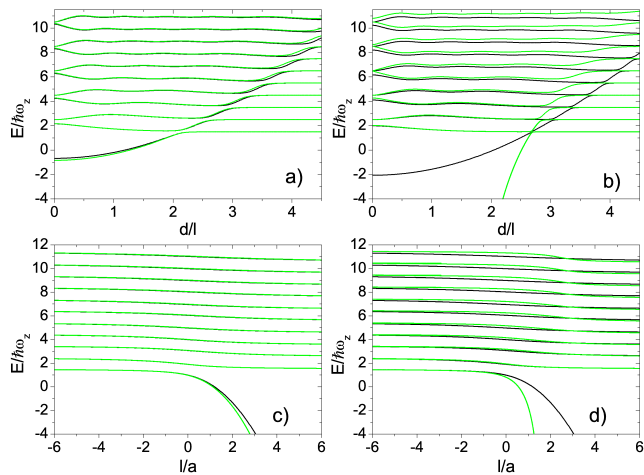}
\caption{
\label{FigOdstepstwa} (Color online)
Upper panels: energy spectrum versus trap separation $d$ for two interacting atoms confined in spherically symmetric trap for $a\approx 0.79 l$. Bottom panels: energy spectrum versus inverse scattering length $1/a$ for atoms in spherically symmetric traps separated by distance $d=1.1 l$.
All lengths are expressed in oscillator units: $l = \sqrt{\hbar/(\mu \omega_z)}$. Black solid curves show the result for the energy-independent pseudopotentials. Green (light grey) solid lines show the energy levels values for the energy-dependent pseudopotential with the phase shift given by the square well potential. The left panels are calculated for a square well with diameter $R_0=0.05 l$ and depth $V=520$ $\hbar \omega$, while the right panels are obtained for $R_0=0.5l$ and $V=520 \hbar \omega$.}
\end{figure}

We illustrate this behavior in Fig.~\ref{FigOdstepstwa}, comparing predictions of the energy-independent and energy-dependent pseudopotentials for a model square-well potential interaction. We note that for a narrow well of the size $R_0=0.05l$, the standard contact potential is quite accurate, with exception of a deep bound state shown in the panel c). In contrast, for a wide square well with $R_0=0.5l$ the two approaches totally differ with respect to the energy dependence of the bound state, and as a consequence the positions of the trap-induced shape resonances shown in panel b). The differences are also visible in the panel d) in the vicinity of the resonance ($a = \pm \infty$), where the standard pseudopotential breaks down due to the lack of the effective range correction.

\subsection{Feshbach resonances}
\label{Sec:Feshbach}

Applying the concept of the energy-dependent scattering length, one can analyze behavior of the system in the presence of Feshbach resonances. To this end we utilize the following result for the energy-dependent scattering length in the vicinity of a magnetic Feshbach resonance \cite{IdziaszekPRA}
\begin{equation}
\label{aFeshbach}
a_\mrm{eff}(E,B)=a_\mrm{bg}\left[1-\frac{\Delta B(1+E/E_b)}{B-(B_0+E/\mu_m-\Delta B E/E_b)}\right].
\end{equation}
Here, $a_\mrm{bg}$ is the background scattering length, $E_b=\hbar^2/(2\mu a_\mrm{bg}^2)$, $\mu_m$ is a difference of magnetic moments between open and closed channels, $B_0$ is a resonance position and $\Delta B$ denotes its width. This expression can be derived from the analytic theory of Feshbach resonances \cite{Tommasini,PaulRMP}, starting from the Breit-Wigner formula for the phase shift near a resonance \cite{IdziaszekPRA}. Eq.~\eqref{aFeshbach} accounts for the effects of the finite kinetic energy of colliding atoms and applies both to the entrance-channel dominated (so called "wide") and closed-channel dominated (so called "narrow") Feshbach resonances. The main correction due to the finite kinetic energy in the open channel is a shift of the resonance $\delta B = E/\mu_m$, which is included in the denominator of Eq.~\eqref{aFeshbach}. The term $\Delta B E/E_b$ represent the correction, which, as we show below, is important for the case of a narrow resonance, whereas it can be safely neglected in the case of a wide resonance. Finally $(1+E/E_b)$ in the numerator of \eqref{aFeshbach} describes the modification of the resonance width at finite energies.

A useful parameter that can be used to classify the resonances is $\eta = (\bar{a}/a_\mrm{bg}) \hbar^2/(2 \mu \bar{a}^2 \Delta B \mu_m)$ \cite{PaulRMP}, where $\bar{a}$ is the mean scattering length $\bar{a} = 0.956 R_\mrm{vdW}$ of the van der Waals potential. For entrance- and closed-channel dominated resonances $\eta \ll 1$ and $\eta \gg 1$, respectively. Since $a_\mrm{bg} \sim \bar{a}$, the above conditions can be rewritten as $\Delta B \mu_m \gg E_b$ and $\Delta B \mu_m \ll E_b$, respectively \footnote{
The fraction $Z$ of the molecular wave function in the closed channel can be obtained from the binding energy of the dressed molecular state $Z =(1/\mu_m) \partial{E_\mrm{mol}(B)}/{\partial B}$ \cite{PaulRMP}, by applying the formula $E_\mrm{mol}(B) = - \hbar^2/(2 \mu a_\mrm{eff}(E_\mrm{mol},B)^2)$ for the energy of the molecule. This yields $Z = 1/\left[1 + \frac12(a_\mrm{eff}(B)/a_\mrm{bg}) \Delta B \mu_m/E_b\right]$, which results in the conditions $\Delta B \mu_m \gg E_b$ and $\Delta B \mu_m \ll E_b$ for entrance- and closed-channel dominated resonances, respectively}.
Hence, for the open-channel dominated resonance one can neglect the term $\Delta B E/E_b$  with respect to $E/\mu_m$ in the denominator of \eqref{aFeshbach}, while for the closed-channel dominated resonances the contribution from $\Delta B E/E_b$ is important.

For weak trapping potentials: $\hbar \omega \ll E_b$ and $\hbar \omega \ll \mu_m \Delta B$, one can omit the corrections due to the finite kinetic energy, and in this case Eq.~\eqref{aFeshbach} reduces to the well-known result $a_\mrm{eff}(B)=a_{bg}\left[1-\Delta B/(B-B_0)\right]$ for the scattering length in the vicinity of a magnetic Feshbach resonance. In general case, however, one can apply the energy-dependent scattering length $a_\mrm{eff}(E,B)$, which properly accounts for the effects of tight trapping potentials \cite{Bolda2003,IdziaszekPRA}.

Here we present some example calculation for a specific system of two ${}^{40}$K atoms. For $s$-wave collisions between hyperfine states $|F=9/2,M_F = -9/2\rangle$ and $|F=9/2,M_F = -5/2\rangle$ we use the following parameters \cite{PaulRMP}: $a_{bg}=174 a_\mrm{Bohr}$, $\Delta B=0.97$~mT, $B_0=22.421$~mT, $\mu_m/h=21.8$~MHz/mT, and $\omega= 2 \pi \times 100$~kHz. For this choice, $\eta =0.4$ and this resonance can be classified as an intermediate between the entrance- and closed-channel dominated limits. Hence, in our calculations we apply the full expression \eqref{aFeshbach} for the energy-dependent scattering length.

\begin{figure}
\includegraphics[width=8.6cm,clip]{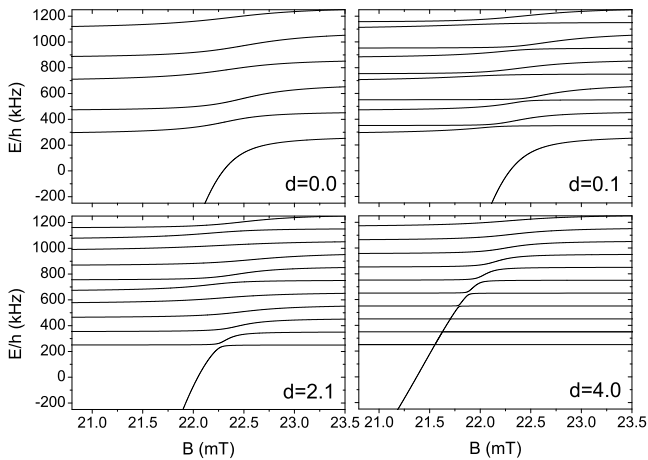}
\caption{
\label{Fig8}
Energy spectrum versus the magnetic field for two ${}^{40}$K atoms in separated harmonic traps in the vicinity of a Feshbach resonance at $B_0=22.42$mT. The results are calculated for $\eta=2$, $\omega=2\pi\times 100$kHz, $d=0.0$, $0.1$, $2.1$, $4.0l$, with $l=\sqrt{\hbar/(\mu \omega_z)}$.}
\end{figure}

\begin{figure}
\includegraphics[width=8.6cm,clip]{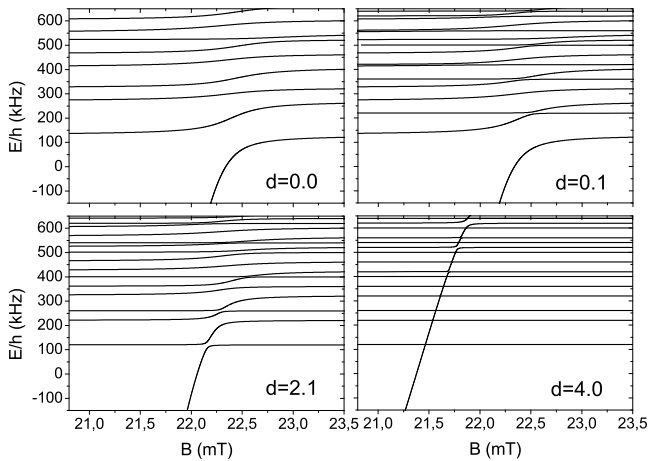}
\caption{
\label{Fig9}
Energy spectrum versus the magnetic field for two ${}^{40}$K atoms in separated harmonic traps in the vicinity of a Feshbach resonance at $B_0=22.42$mT. The results are calculated for $\eta=0.7$, $\omega=2\pi\times 100$kHz, $d=0.0$, $0.1$, $2.1$, $4.0l$, with $l=\sqrt{\hbar/(\mu \omega_z)}$.}
\end{figure}

Figures \ref{Fig8} and \ref{Fig9} show the energy spectrum for the trap aspect ratios $\eta =2$ and $\eta=0.7$, respectively, and for different trap separations. We observe a similar structure of avoided crossings as in Figs.~\ref{Fig4} and \ref{Fig5}.  For $d=0$ where only states with $s$-wave symmetry contribute, the spectrum is regular without avoided crossings. Increasing $d$ results in the appearance of the oscillator states with odd $n_z$, creating avoided crossing with the rest of the states. Finally for sufficiently large $d$, the trap-induced resonances start to appear.


\section{Wave functions}
\label{Sec:Wave}

\subsection{Analytic expressions}

We turn to the analysis of eigenfunctions. For an eigenenergy $\epsilon$ the corresponding wave function $\Psi_{\epsilon}$ is given by Eq.~(\ref{falowa}), or its integral representation \eqref{IntRepr}. In fact those formulas are not very convenient for the numerical evaluation of $\Psi_{\epsilon}(r)$, because the former involves double summation which converges slowly, while the latter is valid only at $\varepsilon <0$. For the purpose of numerical calculations we derive yet another representation containing only a single summation. To this end we express the denominator in Eq.~(\ref{falowa}) in terms of the integral \eqref{tworz}, and we perform only a single summation either over $n$ or $k$ quantum numbers. Summing over $n$ with the help of the generating function (\ref{rownlag}) leads to
\begin{equation}
\begin{split}
&\Psi_{\epsilon}(\mbf{r})=\frac{e^{-\frac{d^2}{2}-\frac{(z-d)^2}{2} -\eta \frac{\rho^2}{2}} }{\pi^{3/2}} \\
&\times\sum_{k} \frac{H_{k}^{*}(-d)H_{k}(z-d)}{2^{k+1} k!}\Gamma(\frac{k-\epsilon}{2 \eta}) U(\frac{k-\epsilon}{2 \eta},1,\eta \rho^2),
\end{split}
\end{equation}
where $U(a,b,x)$ denotes confluent hypergeometric function. Both $\Gamma(x)$ and $U(a,b,x)$ are well defined arbitrary values of the parameters $a$, $b$ and $x$, hence this representation is applicable for all values of the energy. One can easily check that for small $\mathbf{r}$ the wave function is singular and behaves as $1/r$, which is an expected behavior resulting from the application of delta pseudopotential.

When we instead sum over the quantum number $k$ in Eq.~(\ref{falowa}), with the help of (\ref{rownherm}) we obtain
\begin{equation}
\begin{split}
&\Psi_{\epsilon}=\frac{\eta}{\pi^{3/2}} e^{ -\eta \frac{\rho^2}{2}} \\
&\times \sum_{n} L_n(\eta \rho^2)\int_{0}^{\infty}\!\!\!\!dt\, e^{-t(2 \eta n -\epsilon)}\frac{e^{-\frac{d^2+(z-d)^2}{2}\coth(t)+\frac{d(d-z)}{\sinh(t)}}}{\sqrt{1-e^{-2 t}}}.
\end{split}
\end{equation}
In this case the integral cannot be expressed in terms of a closed formula, except the special case of $d=0$ \cite{IdziaszekPRA}.

\subsection{Results}

\begin{figure*}[htbp]
  \begin{center}
    \includegraphics[width=\textwidth]{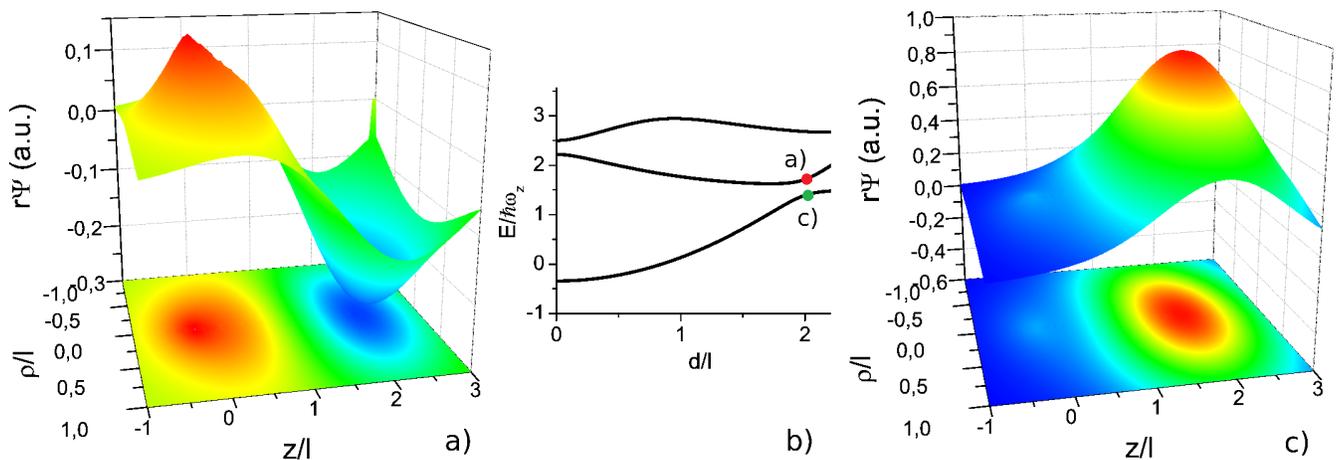}
    \caption{\label{Fig6iFig7} (Color online)
    Panels a) and c): wave function of two interacting atoms confined in isotropic harmonic potentials separated by the distance $d=2.1 l$, for the scattering length $a = l$, and corresponding eigenergies  $E = 0.048 \hbar\omega$ and $E = 0.327 \hbar\omega$, respectively. Panel b): energy spectrum versus trap separation $d$. The red (upper) and the green (lower) dots indicate the position of eigenstates shown in panels a) and c), respectively. All the lengths are expressed in harmonic oscillator units $l = \sqrt{\hbar/(\mu \omega)}$.
}
  \end{center}
\end{figure*}

In Figure~\ref{Fig6iFig7} we present two example wave functions calculated for $a=l$, $\eta=1$, in the vicinity of an avoided crossing at $d=2.1 l$. The corresponding eigenvalues, $E=-0.048 \hbar\omega$ and $E\approx0.327 \hbar\omega$ are shown in the panel b). Both wave functions are multiplied by $r$ in order to obtain a regular behavior at $r \rightarrow 0$. The wave functions presented in panel a) and c) are, respectively, antisymmetric and symmetric combinations of the bound and of the trap states. When, moving away from the trap-induced resonance region, the wave functions
become dominated by the single component localized either at $\mbf{r}= d \hat{\mbf{z}}$ or at $\mbf{r}= 0$, taking the character of trap vibrational or the bound state, respectively.

\section{Example applications of the trap-induced resonances}
\label{Sec:Examples}

To illustrate applicability of our results we consider two example systems, where the trap-induced shape resonances can be applied and can bring some advantages over the standard schemes based on collisions in the same potential well. First we consider the quantum phase gate, involving a pair of separated particles, and controlled by external magnetic field. As a second example we present a scheme allowing for a quantum state control of the relative motion in the trap. In the case of atoms in an optical lattice this method allows for a coherent transfer of particle into higher Bloch bands. In our analysis we do not calculate the full quantum dynamics for the considered schemes, but rather focus on the regime where the dynamics can be described in the adiabatic or diabatic approximation, using the Landau-Zener theory.

\subsection{Quantum gate with two separate atoms}

We consider a relatively simple quantum gate build up of two atoms trapped in the separate trapping potentials. The particles are initially prepared in the ground state of the external motion in the trap, and the qubit states are encoded in the internal spin (hyperfine) states of the atoms. Our scheme assumes that the particles remain at fixed positions during the gate operation, and the state dependent dynamics required for the quantum gate operation is gained by applying an external magnetic field. This differs from the majority of other theoretical proposals when the particles are moved to perform the gate operation, which can reduce the fidelity due to the excitations to higher vibrational states.

We denote the qubit states as $|0\rangle_L$ ($|0\rangle_R$) and $|1\rangle_L$ ($|1\rangle_R$) for particles in the left (right) trap, respectively. The two-qubit states are denoted as $|ij\rangle \equiv |i\rangle_L |j\rangle_R$, for $i,j=0,1$. The goal is to realize the conditional phase gate that is described by the following truth table:
\begin{equation}
\begin{split}
\label{Eq:phaseGate}
|00\rangle &\stackrel{U_{\mathrm{B}}}{\longrightarrow} e^{i\phi_{00}}|00\rangle \stackrel{          U_{\mathrm{S}}  }{\longrightarrow} \phantom{e^{i\phi}}|00\rangle,\\
|01\rangle &\stackrel{\phantom{U_{\mathrm{B}}}}{\longrightarrow} e^{i\phi_{01}}|01\rangle \stackrel{ \phantom{U_{\mathrm{S}}} }{\longrightarrow} \phantom{e^{i\phi}}|01\rangle,\\
|10\rangle &\stackrel{\phantom{U_{\mathrm{B}}}}{\longrightarrow} e^{i\phi_{10}}|10\rangle \stackrel{ \phantom{U_{\mathrm{S}}} }{\longrightarrow} \phantom{e^{i\phi}}|10\rangle,\\
|11\rangle &\stackrel{\phantom{U_{\mathrm{B}}}}{\longrightarrow} e^{i\phi_{11}}|11\rangle
\stackrel{\phantom{U_{\mathrm{S}}} }{\longrightarrow}e^{i\phi} |11\rangle,
\end{split}
\end{equation}
The first transformation $U_\mrm{B}$ is the actual gate process performed with application of an external magnetic field, where $\phi_{ij}$ describe the phases gained by the two-qubit states $|i\rangle_L |j\rangle_R$. The second transformation $U_\mrm{S}$ represent the single qubit operations that can be applied to undo the dynamic phases in all, except the single channel \cite{Calarco2001}, which gains the phase $\phi = \phi_{00}+\phi_{11}-\phi_{10}-\phi_{01}$ \footnote{In the case of unlike atoms, the single qubit operations can be performed by applying appropriate microwave pulses, that can address selectively the atoms in different traps, because of different hyperfine structure of particles. In the case of like species, one can use the fact that the atoms are spatially separated, which allows to address them selectively
\cite{Phillips}.}. For $\phi =\pi$ a so-called phase gate is achieved, which combined with single qubit rotations, form a controlled-NOT. This is particularly interesting since controlled-NOT together with single qubit operations constitute a universal set for quantum computation \cite{Chuang}.

\begin{figure}
\includegraphics[width=8.6cm,clip]{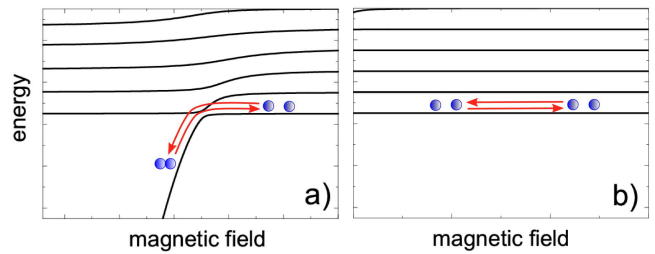}
\caption{
\label{Fig10} (Color online)
Quantum gate for atoms in separate traps controlled by external magnetic fields. Depending on the two-qubit states the atoms would experience (panel a) or not (panel b) the Feshbach resonance, which leads to the state dependent evolution.}
\end{figure}

The main idea behind our quantum gate is to exploit the dependence of the avoided crossings positions on the individual two-qubit states \cite{Hauke}. This dependence stems from the fact that each of the two-qubit states, which can be treated as separate scattering channel, will exhibit distinct structure of the trap-induced shape resonances, resulting from differences in the actual positions of the molecular states (except states $|01\rangle$ and $|10\rangle$ in the case of like atoms).
In this way, Feshbach resonances in each of the channel state appear at different values of the magnetic field, which allows to realize the state-dependent dynamics. In some range of the magnetic field, certain two-qubit states can exhibit an avoided crossings (see Fig~\ref{Fig10}.a), whereas the other states will not experience a resonance (see Fig~\ref{Fig10}.b).

By changing the magnetic field we pass the avoided crossing adiabatically, starting from large toward the small magnetic fields, in order to create molecular states in the two-qubit states for which the Feshbach resonance occur (see Fig~\ref{Fig10}.a). Then we reverse the process and return with the magnetic field to its initial value. At the same time the remaining two-qubit states do not experience the resonance (see Fig~\ref{Fig10}.b) and they only acquire the phase due to the Zeeman shifts of the hyperfine levels.

For simplicity in our analysis we consider the evolution in the adiabatic regime, where the dynamics in the vicinity of an avoided crossing can be described in the framework of the Landau Zener theory.
In the case of an adiabatic process $\phi_{ij} = \int_{t_\mrm{i}}^{t_\mrm{f}} E_{ij}(B(t)) dt$, where the whole operation takes place between $t_\mrm{i}$ and $t_\mrm{f}$ and $E_{ij}(B)=E_i(B)+E_j(B)$ denotes the eigenenergy for a two-qubit state $|ij\rangle$.

The main limitation to the gate fidelity results from the nonadiabatic transitions to other states of external motion in the trap, while passing an avoided crossing. This yields a lower bound for the gate operation time. It can be estimated by applying the Landau-Zener formula for the probability of an adiabatic transfer across the avoided crossing
\begin{equation}
P_\mrm{ad}=1 - \exp\left(-\frac{(\Delta E)^2}{\frac{\partial E}{\partial B} v} \right)
\end{equation}
and assuming the linear ramp of the magnetic field: $v = |\partial B/\partial t| = \mrm{const}$. Here $\Delta E$ denotes the level separation at the avoided crossing, $\partial E/\partial B = \partial E_1/\partial B - \partial E_2/\partial B$ where $E_1(B)$, $E_2(B)$ are the energies of the two levels far from the avoided crossing, and $v=|\partial B/\partial t|$ is a ramp speed. For our example calculation of the Feshbach resonance for ${}^{40}$K atoms (see Fig.~\ref{Fig8}), we estimate that for trap separations $d/l=2.1$ the transfer fidelity $F =0.999$ can be achieved with the speed limited to $v = 8$mT/s, resulting in the total operation around $1$ms. The value of $F =0.999$ is the order of minimal fidelity required to apply the quantum error correction schemes. The gate operation time may be further improved by the optimal control techniques \cite{OptimalControl1,OptimalControl2,Hauke}. Finally we note that the gate operation time will crucially depend on the geometry of the traps, which trivially stems from the fact that for large $\eta$ states are elongated along the symmetry axis, which leads to stronger overlap of the ground-state wave functions in the traps and as a consequence to stronger avoided crossings.

\subsection{Transport to higher bands using trap-induced shape resonances}

As a second example we consider a scheme for a quantum state control of a pair of atoms in a harmonic trap. It allows for a transfer of particles between arbitrary states of the relative motion in a trap. A similar method, based on the use of Feshbach resonances has been applied experimentally, to transfer a mixture of fermions trapped in a deep optical lattice into higher Bloch bands \cite{Kohl}. In our scheme we combine the technique of a magnetic field sweep with the use of trap-induced shape resonances, to obtain a method for an efficient transfer of atoms into arbitrary excited state of the relative motion. This can be applied, for instance, to probe states of atoms in higher Bloch bands, using the similar setup as in Ref. \cite{Kohl}. This requires that external potential for different spin states can be controlled separately, e.g. by means of spin-dependent lattice potentials.

\begin{figure}
\includegraphics[width=8.6cm,clip]{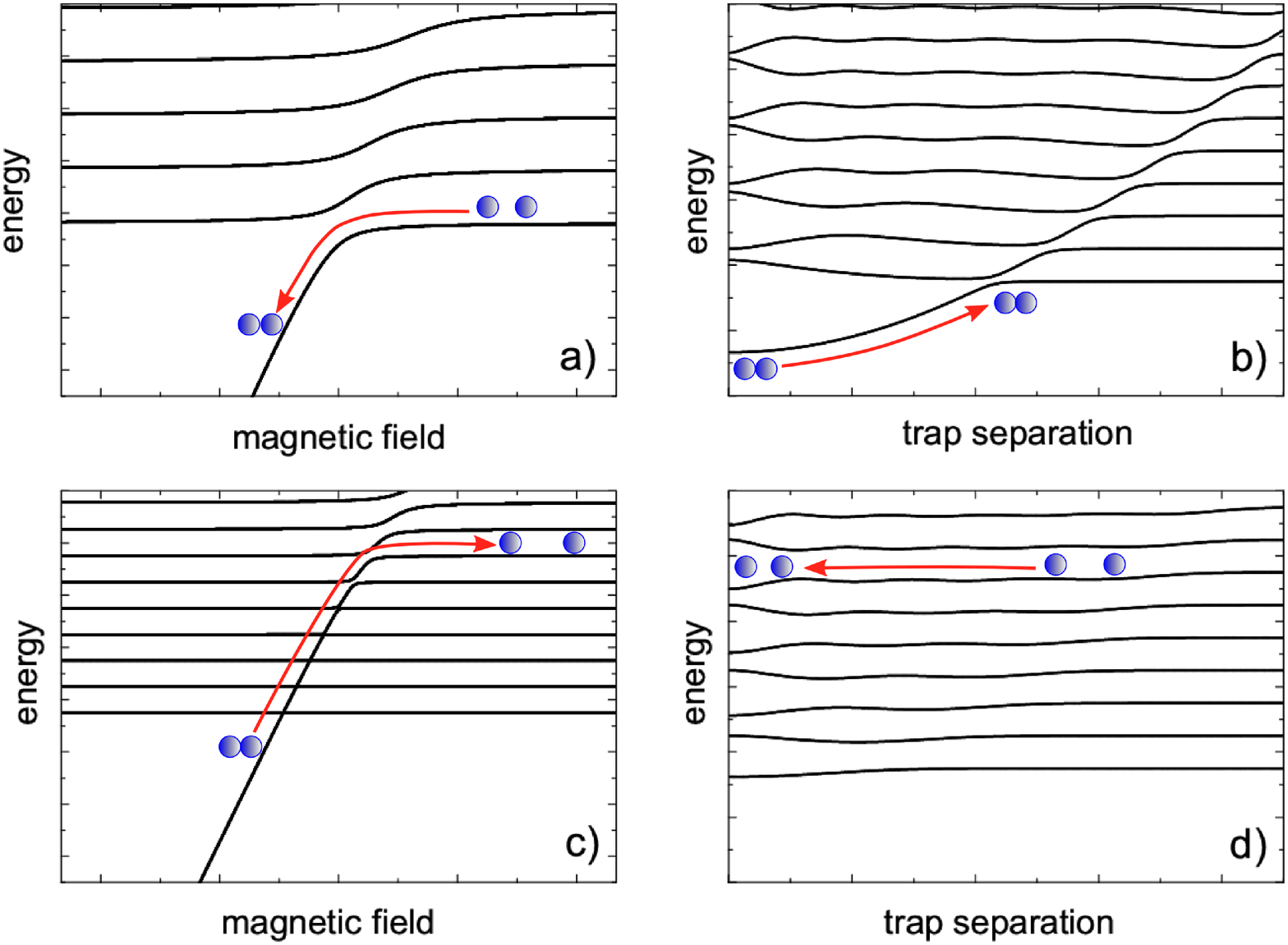}
\caption{
\label{Fig11} (Color online)
Coherent transfer of a pair of atoms into an excited state: a) conversion of pair of atoms into molecules by a magnetic field sweep; b) transfer of a molecular state to the region of an avoided crossing (trap-induced shape resonance) by changing trap separation; c) conversion of a molecule into an excited state in the trap by a magnetic field sweep; d) transferring the state back to the regime of a zero trap separation. We end with two particles in an excited state of $s$ or $p$-wave symmetry.}
\end{figure}

We assume that initially the atoms are prepared in the ground state of the trap, and the magnetic field is set at the attractive side of a Feshbach resonance. Our scheme consists of the following steps (see Fig.~\ref{Fig11}): a) magnetic field sweep across Feshbach resonance resulting in an adiabatic conversion of a pair of atoms into a molecule; b) increasing the energy of a molecule by changing the trap separation from 0 to some finite value $d$, before reaching the avoided crossing
c) conversion of molecule into an excited state of atoms in the trap, using the sweep across Feshbach resonance d) bringing back the traps to an initial positions ($d =0$). Steps a) and c) assume constant trap separation, whereas b) and d) assume constant value of the magnetic field. In principle one could exchange the order of steps b) and c), converting a molecule directly into excited state of the trap, and later change the value of the magnetic field from repulsive
to attractive interactions. Fig.~\ref{Fig11} shows, however, that such a scenario could be more difficult to realize, since the strength of avoided crossing is of the same order for low-lying and highly excited states, when changing the trap distance $d$ instead of a magnetic field (compare panels b and c)

Using the Landau-Zener theory, we made an estimation for an adiabatic transfer from the vibrational ground-state to the third excited state of a relative motion. We took parameters for two
${}^{40}$K atoms, near a Feshbach resonance at $B_0=22.421$~mT, assuming spherically symmetric trap ($\eta=1$). The main limitation for the rate of transfer results from the dynamics during the stage c), where the first two avoided crossings have to be traversed diabatically, whereas the third one adiabatically. We investigated the simplest case of a constant sweep rate across the Feshbach resonance. For a sweep at rate $v=|\partial B/\partial t|\approx 0.2$mT/s, the multiple-crossing Landau-Zener theory \cite{Multi1,Multi2}  predicts that atoms at the stage c) can be transferred from the molecular to the third excited state with efficiency greater than $75\%$.
In contrast at stage a), the avoided crossing is much stronger and, for instance, conversion with probability $99\%$ is obtained with much faster rate $v=0.35$T/s.
Steps b) and d) can be also performed relatively fast, with speed limited by the time scale given by the level separation in the trap $\dot{d} \ll l \omega$, with $l = \sqrt{\hbar/(\mu \omega)}$. Finally all the steps can be further optimized applying the optimal control techniques, which should significantly reduce the time of the whole process. For instance a transfer of a particle in a harmonic trap to a distance of the order of few harmonic oscillator lengths can be performed on the scale of a single trap period $T=2 \pi/\omega$, by appropriate optimizing the transfer process \cite{Tommaso}.

\section{Summary}
\label{Sec:Conclusions}

In this paper we have presented analytical solutions for the system of two atoms trapped in harmonic potentials separated by an arbitrary distance $d$. The atom-atom interaction have been modeled with the Fermi pseudopotential, including the energy-dependent scattering length. This accounts for the presence of tight traps and scattering resonances. We have investigated the properties of the energy spectrum and of the wave functions for different trap separations, scattering lengths, and aspect ratios of the cylindrically symmetric trap. The spectrum exhibit avoided crossings resulting from the resonances between the molecular and trap states. For isotropic traps ($\eta=1$) the trap vibrational states at large $d$, correlate at small $d$ with the states of two interacting atoms with $s$ and $p$-wave symmetry. One can explain this behavior in the framework of the perturbation theory.
Finally we illustrated our analytical results with two physical examples, where the trap-induced shape resonances can be applied. The first one was a quantum phase gate for two separated particles, based on a control of the magnetic field. The second one was a scheme for a coherent transport of particles into higher Bloch bands.

In our work we have not considered the possibility of the two different trapping frequencies for interacting atoms. This can be particularly important for atoms of different species or in the case of the ion-atom collisions. The inclusion of the two different trapping frequencies can lead to the additional resonances between center-of-mass and relative motions, that should be present already at zero trap separations.


\begin{acknowledgments}
The authors are grateful to A. Witkowska and H. Doerk-Bendig for valuable discussions. This work was supported by the Polish Government Research Grant for years 2007-2010.
\end{acknowledgments}

\appendix

\section{Generating functions for harmonic oscillator states}
\label{App:Sums}

The harmonic oscillator functions in one and two-dimensional traps are given by, respectively,
\begin{equation}
  \Phi_{n,0}(\rho,\varphi)=\frac{\sqrt{\eta}}{\sqrt{\pi}} e^{-\eta \frac{\rho^2}{2}}L_n(\eta\rho^2), \quad m=0,
  \label{phin0}
\end{equation}
\begin{equation}
  \Theta_k(z)=\frac{e^{-\frac{z^2}{2}}}{\pi^{\frac{1}{4}} \sqrt{2^k k!}}H_k(z).
  \label{thetak}
\end{equation}
In derivation of the integral representation of the wave function, we have applied the generating functions for Laguerre
\begin{equation}
  \sum_{n=0}^{\infty}L_n(x)z^n=(1-z)^{-1}\exp(\frac{x z}{z-1}),
  \label{rownlag}
\end{equation}
and Hermite polynomials
\begin{equation}
  \sum_{k=0}^{\infty}\frac{t^k}{2^k k!}H_k(x)H_k(y)=\frac{e^\frac{2 t x y-t^2 x^2-t^2 y^2}{(1-t)^2}}{\sqrt{1-t^2}},
  \label{rownherm}
\end{equation}

\section{Derivation of the recurrence relations}
\label{App:Recur}

Let's consider the difference between values of ${\cal F}(x)$ separated by $\eta$
\begin{equation}
{\cal F}(x)-{\cal F}(x+\eta)=\int_0^{\infty}dt{\frac{\eta e^{-x t} e^{-d^2 \frac{1-e^{-t/2}}{1+e^{-t/2}}}}{\sqrt{1-e^{-t}}}}.
\end{equation}
After the substitution of $y=\tanh(\frac{t}{4})$ this can be transformed into the following form
\begin{equation}
   {\cal F}(x)-{\cal F}(x+\eta)=2 \eta\!\!\int_0^{1}\!\!\!dy\,e^{-d^2 y} y^{-1/2}(1-y)^{2x-1}(1+y)^{-2x}.
\label{roznica f}
\end{equation}
The integral on the r.h.s. can be evaluated analytically with the help of \cite{Gradshteyn}
\begin{multline}
  \int_{0}^{1}dx x^{\nu-1}(1-x)^{\lambda-1}(1-\beta x)^{-\rho}e^{-\mu x}\\=B(\nu,\lambda)\Phi_1 (\nu,\rho,\lambda+\nu;\beta;-\mu)
\label{calka}
\end{multline}
valid for $\mrm{Re}(\lambda) >0$, $\mrm{Re}(\nu) >0$, $|{\arg(1-\beta)}|<\pi $. This yields
\begin{equation}
{\cal F}(x)-{\cal F}(x+\eta)=B\left(\textstyle{\frac12},2x\right) \Phi_1\left(\textstyle{\frac12},2x,2x+\textstyle{\frac12};-1,-d^2\right),
\label{Recur2}
\end{equation}
Here, $\Phi_1(a,b,c,x,y)$ is the confluent hypergeometric function of two variables, which can be defined by the series \cite{Gradshteyn}
\begin{equation}
  \Phi_1(a,b,c,x,y)=\sum_{m,n=0}^{\infty}\frac{(a)_{m+n} (b)_n}{(c)_{m+n} m! n!}x^m y^n.
\end{equation}
Evaluation of $\Phi_1(a,b,c,x,y)$ in terms of a double series is not convenient, and in the numerical calculations we have applied another result, which can be derived by applying the binomial expansion
\begin{equation}
(1+x)^a=\sum_{k=0}^{\infty}\frac{(-x)^k \Gamma(k-a)}{k! \Gamma(-a)}
\end{equation}
to the terms $(1-y)^{2x-1}$ or $(1+y)^{-2x}$ in \eqref{roznica f}. In this case the remaining integral can be calculated with the help of \cite{Gradshteyn}
\begin{multline}
\int_{0}^{u} dx x^{\nu-1}(u-x)^{\mu-1}e^{\beta x} \\
=B(\mu,\nu)u^{\mu+\nu-1}{{}_1 F_1}(\nu;\mu+\nu;\beta u), \quad \mu>0, \nu>0
\end{multline}
In the way described above we obtain two series expansions for ${\cal F}(x)-{\cal F}(x+\eta)$:
\begin{equation}
{\cal F}(x)-{\cal F}(x+\eta)=\sum_{n=0}^{\infty} a_n
\label{eq4}
\end{equation}
with
\begin{align}
a_n & =\sum_{n=0}^{\infty} \Big[ (-1)^n \frac{\Gamma(n+2x)}{\Gamma(2x) n!}  \nonumber \\
& \phantom{=} \times B(2x,n+\tst{\frac12}){{}_1 F_1}(n+\tst{\frac12};2x+n+\tst{\frac12};-d^2)\Big],
\end{align}
and
\begin{equation}
{\cal F}(x)-{\cal F}(x+\eta) = \frac{2^{2x-1}}{\Gamma(2x)}\sum_{n=0}^{\infty} b_n
\label{eq1}
\end{equation}
with
\begin{equation}
   b_n = 2^n\frac{(\Gamma(n+2x))^3}{n! \Gamma(2n+4x)}{{}_1 F_1}(\tst{\frac{1}{2}};2x+n+\tst{\frac{1}{2}};-d^2)
\end{equation}
For large $n$, the first series \eqref{eq4} converges rather slowly:
$a_n+a_{n+1} \sim \exp(-d^2)/n^2$ for $n$ odd. In numerical calculations it is easier to apply the second series \eqref{eq1}, which has the asymptotic behavior given by: $b_n \sim \sqrt{2\pi}(n+2x)^{2x-3/2}/2^{n+4x-1/2}$.

\bibliography{contrcoll}

\end{document}